\documentclass[a4paper,fleqn]{cas-sc}

\usepackage[numbers,sort&compress]{natbib}
\usepackage{gensymb}
\usepackage{tablefootnote}
\usepackage{refcount}
\usepackage{threeparttable}
\usepackage{hyperref}
\def\tsc#1{\csdef{#1}{\textsc{\lowercase{#1}}\xspace}}
\tsc{WGM}
\tsc{QE}
\begin{document}
\let\WriteBookmarks\relax
\def\floatpagepagefraction{1}
\def\textpagefraction{.001}

\shorttitle{Analysis of \textsuperscript{131}Xe Prompt Gammas}    

\shortauthors{A. Rosen, J. Moldenhauer, C. Michelagnoli, M. Stryjczyk, W. Flanagan}   %C. Michelagnoli

\title [mode = title]{Analysis of \textsuperscript{131}Xe Prompt Gammas from Thermal Neutron Capture at FIPPS}  




\author[1]{Andrew Rosen}[orcid=0009-0007-3975-5442]

%\cormark[1]


\ead{adr4278@my.utexas.edu}



\author[2]{Jacob Moldenhauer}%[]

% Footnote of the second author

% Email id of the second author



%\author[3]{Caterina Michelagnoli}%[]

% Footnote of the second author

% Email id of the second author



\author[3]{Caterina Michelagnoli}

\author[3]{Marek Stryjczyk}%[]

% Footnote of the second author

% Email id of the second author


\author[1,2]{Will Flanagan}%[]
\ead{will.flanagan@austin.utexas.edu}
% Footnote of the second author

% Email id of the second author



\affiliation[1]{organization={The University of Texas at Austin},
            addressline={11131 Creativity Trail}, 
            city={Austin},
%          citysep={}, % Uncomment if no comma needed between city and postcode
            postcode={78758}, 
            state={Texas},
            country={USA}}
\affiliation[2]{organization={University of Dallas},
            addressline={1845 E Northgate Dr}, 
            city={Irving},
%          citysep={}, % Uncomment if no comma needed between city and postcode
            postcode={75062}, 
            state={Texas},
            country={USA}}
\affiliation[3]{organization={Institut Laue-Langevin},
            addressline={, 71 avenue des Martyrs}, 
            city={Grenoble},
%          citysep={}, % Uncomment if no comma needed between city and postcode
            postcode={38000}, 
            %state={},
            country={France}}

%\cortext[1]{Corresponding author}



\begin{abstract}
The prompt gammas of $^{131}$Xe from thermal neutron capture were measured for the first time at the Fission Product Prompt $\gamma$-ray Spectrometer by irradiation of isotopically enriched $^{130}$Xe in a high-purity quartz ampoule. Through $\gamma\gamma$-coincidences, 42 new prompt gammas were measured and 14 new nuclear states determined. %\nocite{*}%% Remove this line from your manuscript. 
\end{abstract}


\begin{highlights}
\item This is the first measurement of thermal neutron capture on an isotopically enriched \textsuperscript{130}Xe target.
\item Through $\gamma$$\gamma$ coincidences, 42 new prompt gammas were measured and 14 new nuclear states determined.
\item All previously measured transitions of \textsuperscript{130}Xe$(\text{n},\gamma)$\textsuperscript{131}Xe are confirmed. 12 primary transitions from the capture state are measured where none were previously known.
\end{highlights}

% Keywords

\maketitle{}

% Main text
\section{Introduction}\label{Introduction}
The prompt $\gamma$-rays of \textsuperscript{131}Xe from thermal neutron capture were measured by irradiation of isotopically enriched \textsuperscript{130}Xe in a high-purity quartz ampoule at the Fission Product Prompt $\gamma$-ray spectrometer (FIPPS) instrument at Institut Laue-Langevin (ILL). Prior to this work, 66 transitions and and 57 excited states were known \cite{ensdf_2026}; 13 prompt gammas for this isotope were tabulated by the IAEA Database for Prompt Gamma-ray Neutron Activation Analysis \cite{iaea_xe130}. These data were collected at the Budapest Research Reactor with a \textsuperscript{nat}XeF\textsubscript{2} target and given isotopic assignments by Rick Firestone based on other existing nuclear data \cite{revay2004}.

\section{Materials and Methods}\label{Mats&Meth}
\begin{figure}%[]
  \centering
    \includegraphics[width=1\linewidth]{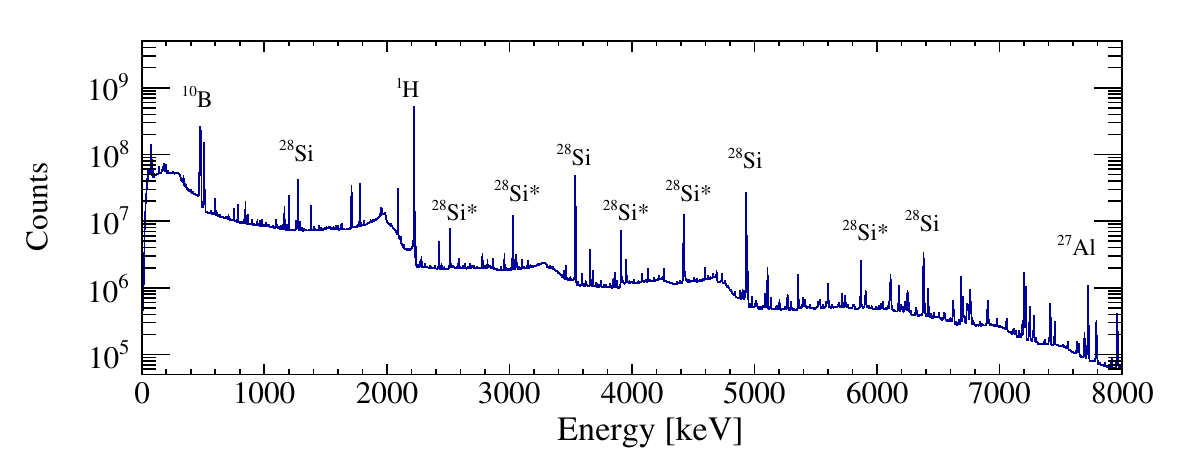}
    \caption{The full spectrum of this work. The most prominent lines are from \textsuperscript{28}Si$(\text{n}, \gamma)$\textsuperscript{29}Si and \textsuperscript{28}Al$(\text{n},\gamma)$\textsuperscript{28}Al from the quartz vial and FIPPS structure, respectively. Escape peaks are denoted by a $\ast$.}\label{FullSpec}
\end{figure}

\begin{figure}%[]
  \centering
    \includegraphics[width=0.5\linewidth]{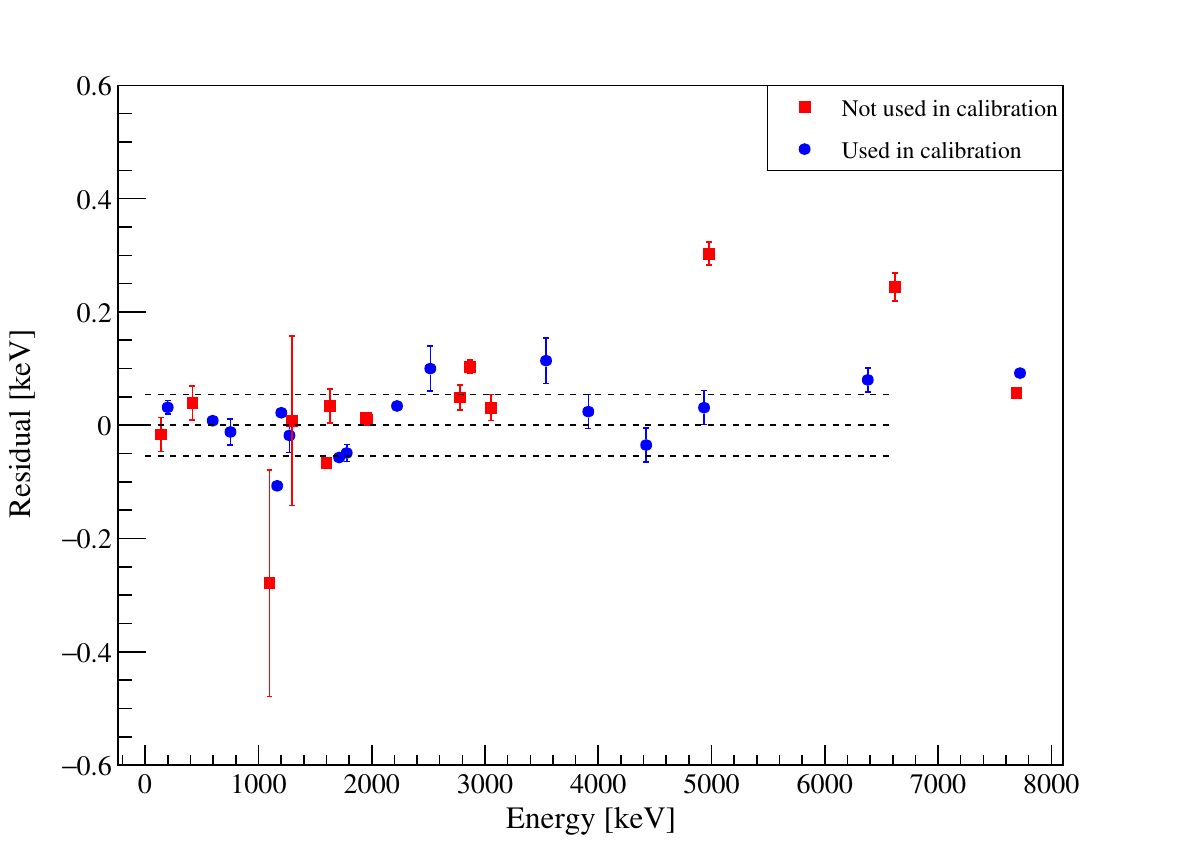}
    \caption{The residuals of the energy calibration. See Table \ref{CalibEnergies_Table} for exact values. The dashed line extends to the \textsuperscript{131}Xe neutron separation energy.}\label{fig1}
\end{figure}

\begin{figure}%[]
  \centering
    \includegraphics[width=0.5\linewidth]{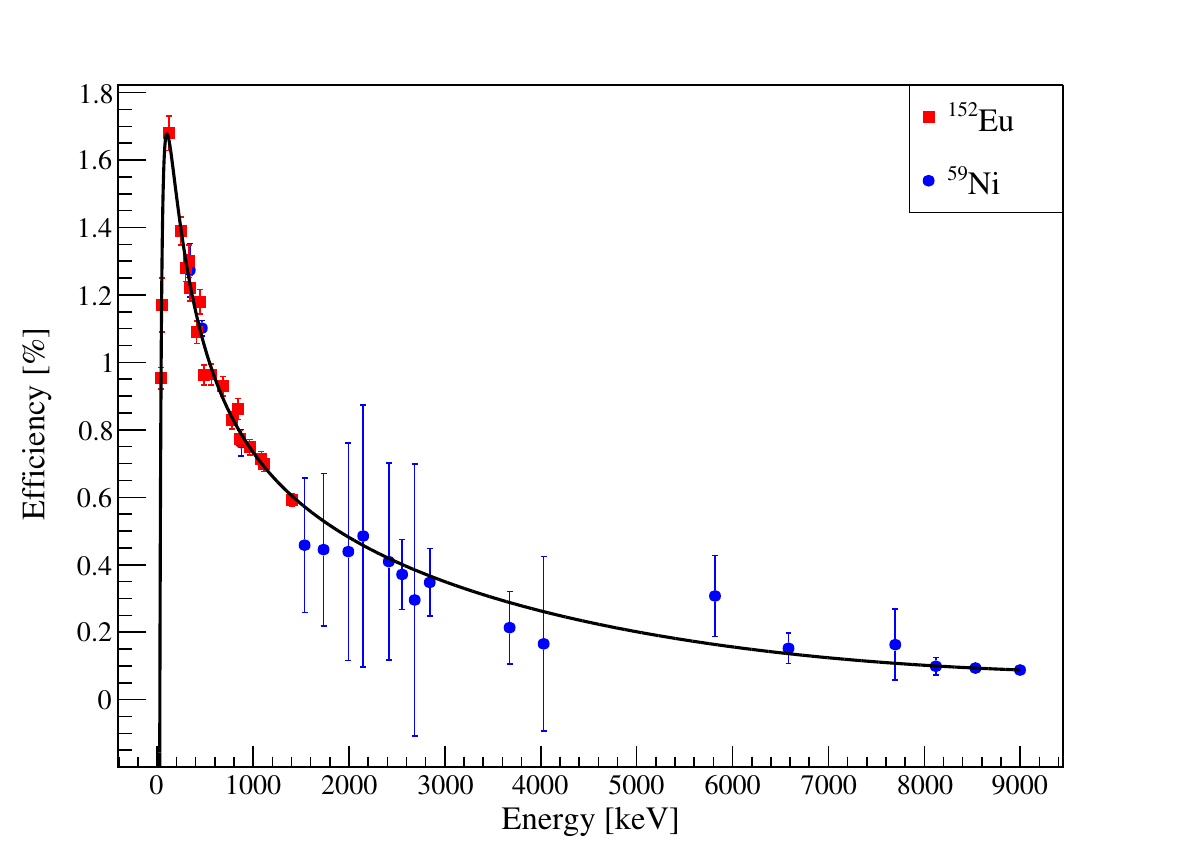}
    \caption{The total efficiency of the FIPPS setup as determined by \textsuperscript{152}Eu and \textsuperscript{59}Ni.}\label{EffGraph}
\end{figure}

The $^{131}$Xe nucleus was studied by neutron capture on
an isotopically enriched $^{130}$Xe target, at the FIssion Product Prompt
$\gamma-$ray Spectrometer (FIPPS) instrument for $\gamma$-ray detection at the Institut Laue-Langevin (ILL) in
Grenoble, France.
ILL hosts a research reactor which generally operates at a thermal power
of 58 MW ~\cite{ill} to produce the most intense continuous neutron flux in the world, namely $1.5\times
10^{15}$ neutrons per second and per cm$^2$; its core is made of a single highly enriched
uranium fuel element that is cooled by heavy water.
The neutrons are transferred to the experimental areas through beam
guides. The beam of the FIPPS instrument is collimated by a sequence of five circular apertures made of boron carbide followed by
5\,cm thick lead absorbers. 
\\
\indent Close to the germanium detectors, $^{6}$LiF
enriched plastic is used for shielding instead of boron to minimize the neutron-induced gamma background. The entire
system is under primary vacuum (1\,mbar) to minimize
neutron scattering in air. The thermal beam at the target position had a
flux of $1 \times 10^8$ n cm$^{-2}$ s$^{-1}$ and a diameter of approximately
1 cm, measured via activation of a gold
foil~\cite{Michelagnoli:2018mgb}. The beam was attenuated using a
$^{6}$LiF-enriched diaphragm because of count-rate capabilities of the data acquisition system at the end of collimation system.
The HPGe-Compton-suppressed clover detector system consisted of 26 HPGe crystals in total: 8 were positioned in the central ring placed at angles of 0\degree  and 135\degree, and 18 were placed at different angles off of the central ring.\footnote{The system was smaller than
the usual array because of maintenance of some of the detectors.}
\\
\indent The target ampoule was placed in the beamline for 86 hours. The most prominent backgrounds were silicon from the sample and aluminum from the setup as shown in Figure \ref{FullSpec}. Due to these large backgrounds, the data calibration was performed primarily with these gammas and their single and double escape peaks (see Table \ref{CalibEnergies_Table} for exact values). The energy uncertainty was determined by the standard deviation of the residuals below the q-value of 6604.419 keV \cite{nndc_qcalc}; this value was found to be 0.054 keV. These residuals and the energy uncertainty are shown in Figure \ref{fig1}.
\\
\indent To determine detector efficiency \cite{biegalski2005}\cite{molnar2002}, a foil of naturally occurring nickel was irradiated in combination with a \textsuperscript{152}Eu source with activity known to 3\%. 
The areas of the \textsuperscript{152}Eu peaks were normalized for intensity and activity; the in-neutron beam nickel data were normalized relative to the 339 keV gamma. The efficiency curve and corresponding data is reported in Figure \ref{EffGraph} and fit using the phenomenological function
\begin{equation}
    \epsilon(E)=\sum_{k=0}^{5}a_{k}\ln^{k}(E)
\end{equation}
whose coefficients are given in Table \ref{EffCoeffs} from the counts given in Table \ref{EffEnergies}. The setup was determined to have a total absolute efficiency of 0.605(11)\% at 1.4 MeV.

\begin{table}
\begin{threeparttable}
\caption{Gammas used for energy calibration and uncertainties with residuals.}
\label{CalibEnergies_Table}
\begin{tabular*}{\tblwidth}{@{}LLLLLL@{}}
\toprule $E_\gamma$ (NNDC)& Reaction (used in calibration) & Residual & $E_\gamma$ (IAEA) & Reaction (not used in calibration) & Residual\tnote{1} \\ 
\midrule
198.371  & 0.032(12)  & {\textsuperscript{71m}Ge(IT)\textsuperscript{71}Ge}         & 139.68   & -0.016(3)  & \textsuperscript{75m}Ge(IT)\textsuperscript{75}Ge           \\
595.851  & 0.008(5)   & {\textsuperscript{73}Ge(n,$\gamma$)\textsuperscript{74}Ge}                   & 416.86   & 0.039(3)   & \textsuperscript{116}In($\beta^{-}$)\textsuperscript{116}Sn \\
752.215  & -0.012(23) & {\textsuperscript{30}Si(n,$\gamma$)\textsuperscript{31}Si}       & 1097.3   & -0.28(20)  & \textsuperscript{116}In($\beta^{-}$)\textsuperscript{116}Sn \\
1164.865 & -0.107(1)  & {\textsuperscript{35}Cl(n,$\gamma$)\textsuperscript{36}Cl}  & 1293.54  & 0.0080(15) & \textsuperscript{116}In($\beta^{-}$)\textsuperscript{116}Sn \\
1201.245 & 0.022(1)   & {\textsuperscript{1}H(n,$\gamma$)\textsuperscript{2}H}                 & 1601.072 & -0.067(4)  & \textsuperscript{35}Cl(n,$\gamma$)\textsuperscript{36}Cl    \\
1273.33  & -0.02(3)   & {\textsuperscript{28}Si(n,$\gamma$)\textsuperscript{29}Si}  & 1633.53  & 0.034(30)  & \textsuperscript{19}F($\beta^{-}$)\textsuperscript{20}Ne    \\
1712.245 & -0.057(1)  & {\textsuperscript{1}H(n,$\gamma$)\textsuperscript{2}H}                 & 1951.14  & 0.013(22)  & \textsuperscript{35}Cl(n,$\gamma$)\textsuperscript{36}Cl    \\
1778.987 & -0.049(15) & {\textsuperscript{28}Al($\beta^{-}$)\textsuperscript{28}Si} & 1959.346 & 0.0090(41) & \textsuperscript{35}Cl(n,$\gamma$)\textsuperscript{36}Cl    \\
2223.245 & 0.034(1)   & {\textsuperscript{1}H(n,$\gamma$)\textsuperscript{2}H}                & 2780.552 & 0.049(22)  & \textsuperscript{30}Si(n,$\gamma$)\textsuperscript{31}Si         \\
2516.92  & 0.10(4)    & {\textsuperscript{28}Si(n,$\gamma$)\textsuperscript{29}Si}  & 2863.819 & 0.10(12)   & \textsuperscript{35}Cl(n,$\gamma$)\textsuperscript{36}Cl    \\
3538.92  & 0.11(4)    & {\textsuperscript{28}Si(n,$\gamma$)\textsuperscript{29}Si}  & 3054.321 & 0.031(23)  & \textsuperscript{30}Si(n,$\gamma$)\textsuperscript{31}Si         \\
3911.98  & 0.02(3)    & {\textsuperscript{28}Si(n,$\gamma$)\textsuperscript{29}Si}  & 4979.759 & 0.30(20)   & \textsuperscript{35}Cl(n,$\gamma$)\textsuperscript{36}Cl    \\
4422.98  & -0.04(3)   & {\textsuperscript{28}Si(n,$\gamma$)\textsuperscript{29}Si}  & 6619.615 & 0.24(25)   & \textsuperscript{35}Cl(n,$\gamma$)\textsuperscript{36}Cl    \\
4933.98  & 0.03(3)    & {\textsuperscript{28}Si(n,$\gamma$)\textsuperscript{29}Si}  & 7693.397 & 0.057(72)  & \textsuperscript{27}Al(n,$\gamma$)\textsuperscript{28}Al    \\
6379.801 & 0.080(21)  & {\textsuperscript{27}Al(n,$\gamma$)\textsuperscript{28}Al}             &            &                                   \\
7724.034 & 0.092(41)  & {\textsuperscript{27}Al(n,$\gamma$)\textsuperscript{28}Al}             &            &                                   \\
\bottomrule
\end{tabular*}
\begin{tablenotes}
    \item[1] Given uncertainties are from both the energy of the known line and measured value.
    \item[2] Escape peak
\end{tablenotes}
\end{threeparttable}
\end{table}

\begin{table}
\caption{Coefficients for the efficiency function.}\label{EffCoeffs}
\begin{tabular*}{\tblwidth}{@{}LLLLLL@{}}
\toprule a\textsubscript{0}        & a\textsubscript{1}       & a\textsubscript{2}         & a\textsubscript{3}        & a\textsubscript{4}           & a\textsubscript{5}           \\ 
\midrule
-0.68(13) & 0.51(11) & -0.148(35) & 0.021(55) & -0.00146(42) & 0.000041(13) \\
\bottomrule
\end{tabular*}
\end{table}

\begin{table}
\caption{Relative and absolute efficiencies of the \textsuperscript{152}Eu source and nickel foil.}
\label{EffEnergies}
\begin{tabular*}{\tblwidth}{@{}LLLLLL@{}}
\toprule $E_\gamma$ (\textsuperscript{152}Eu decay)& Absolute Efficiency (\%) & $I_\gamma$ & $E_\gamma$ (\textsuperscript{58}Ni(n,$\gamma$)\textsuperscript{59}Ni) & Relative Efficiency (\%) & $\sigma$ (b) \\ 
\midrule
39.80616 & 0.953(32) & 58.7(9)    & 339.42   & 100.00(8)  & 0.167(10)  \\
45.474   & 1.170(79) & 13.41(82)  & 464.978  & 86.49(2)   & 0.843(10)  \\
121.7474 & 1.682(51) & 28.53(16)  & 877.977  & 59.80(4)   & 0.236(3)   \\
244.6974 & 1.394(42) & 7.550(4)   & 1536.92  & 35.96(20)  & 0.0194(5)  \\
295.938  & 1.283(40) & 0.440(4)   & 1734.687 & 34.93(23)  & 0.0172(4)  \\
329.41   & 1.301(48) & 0.1213(24) & 1949.911 & 58.77(11)  & 0.0476(10) \\
344.2894 & 1.223(38) & 26.6(2)    & 1992.794 & 34.46(32)  & 0.0132(5)  \\
411.1184 & 1.092(33) & 2.237(13)  & 2015.576 & 12.42(52)  & 0.0117(4)  \\
443.9723 & 1.180(36) & 2.827(14)  & 2147.74  & 38.09(39)  & 0.0119(5)  \\
488.6761 & 0.962(30) & 0.414(3)   & 2414.872 & 32.14(29)  & 0.0121(5)  \\
563.9672 & 0.964(31) & 0.494(5)   & 2554.116 & 29.15(10)  & 0.0431(9)  \\
688.6733 & 0.929(29) & 0.856(6)   & 2685.06  & 23.20(40)  & 0.0108(5)  \\
778.8678 & 0.828(25) & 12.93(8)   & 2842.13  & 27.29(10)  & 0.0463(10) \\
841.5038 & 0.862(31) & 0.168(3)   & 3675.24  & 16.76(11)  & 0.0281(7)  \\
867.3339 & 0.774(24) & 4.23(3)    & 4030.14  & 12.99(26)  & 0.0145(5)  \\
964.0297 & 0.749(23) & 14.51(7)   & 5817.219 & 24.13(12)  & 0.1090(22) \\
1085.829 & 0.714(22) & 10.11(5)   & 6583.831 & 11.947(45) & 0.083(2)   \\
1112.08  & 0.698(21) & 13.67(8)   & 7697.163 & 12.798(11) & 0.0374(14) \\
1408.025 & 0.592(18) & 20.87(9)   & 8120.567 & 7.766(26)  & 0.133(3)   \\
         &           &            & 8533.509 & 7.3470(70) & 0.721(13)  \\
         &           &            & 8998.414 & 6.8892(41) & 1.49(3)    \\ \hline
\bottomrule
\end{tabular*}
\end{table}

\section{Analysis}\label{Analysis}
The level structure of \textsuperscript{131}Xe has been previously probed by 7 other techniques \cite{ensdf_2026}. The measurement of \textsuperscript{131}I $\beta{}^{-}$ \cite{meyer1974} gave levels below the $Q_{\beta^-}$ of 970.8 keV \cite{nndc_qcalc} but was not sensitive to higher energy states. Bombardment of tellurium with alpha particles $\left( {}^{128}\text{Te}(\alpha,\text{n}) \text{ and} {}^{130}\text{Te}(\alpha,3\text{n})\right)$\cite{wirth2003} observed nearly all known low-lying states seen with \textsuperscript{131}I $\beta^{-}$  and was additionally sensitive to high spin ($\text{J}^{\pi}$ > 11/2) states above 723 keV. Coulomb excitation \cite{palmer1978} of the nucleus revealed two levels within the low-energy regime that iodine beta decay and tellurium alpha bombardment were not sensitive to. \textsuperscript{131}Xe($\gamma$, $\gamma'$)\textsuperscript{131}Xe \cite{tam1990}\cite{vongarrel2006} is the only previous look into the low-spin (1/2,3/2,5/2) high energy structure of this nucleus. Since the low-energy region has been thoroughly studied, this work seeks to probe low-spin, high excitation energy states above \~700 keV with greater precision, making this work highly complementary to previous measurements.
Of the 57 previously known nuclear states of \textsuperscript{131}Xe, 39 were measured via alpha particle bombardment of \textsuperscript{128}Te and \textsuperscript{130}Te whose spins are significantly greater than the 1/2+ capture state of \textsuperscript{131}Xe, and thus transitions to states with spin greater than 11/2- or 7/2+ were not observed \cite{lonnroth1983}. Of the remaining 19 low-spin states, 17 were observed with three new transitions between known states being found by this work. From these, 28 gammas were observed out of the known 39 - those not observed had relative intensities of about 1\% or less. All 13 previous known prompt gammas from the IAEA Database for Prompt Gamma-ray Neutron Activation Analysis were observed (Table \ref{IAEAGammas}) \cite{iaea_xe130}.
%Many of the new energy levels found by this work are close to those reported by \textsuperscript{131}Xe($\gamma$,$\gamma'$)\textsuperscript{131}Xe, however this classification is merely qualitative due to unknown uncertainties of the ($\gamma$,$\gamma'$) measurement and no observation of direct transitions to the ground state in this work as reported in \cite{vongarrel2006}.
Several of the new energy levels found by this work are close to those reported by a \textsuperscript{131}Xe($\gamma$,$\gamma'$)\textsuperscript{131}Xe measurement, however this classification is merely qualitative due to unknown energy resolution and branching fractions of the excited states reported in that work \cite{vongarrel2006}.

\begin{table}[H!]
\caption{Prompt gammas of \textsuperscript{131}Xe from the IAEA Database for Prompt Gamma-ray Neutron Activation Analysis compared to the values for measured gammas and those from the NNDC. All energies are given in keV.}\label{IAEAGammas}
\begin{tabular*}{\tblwidth}{@{}LLL@{}}
\toprule $E_\gamma$ (IAEA) & $E_\gamma$ (NNDC)& $E_\gamma$ (This work)\\ 
\midrule
295.30(20) & 295.3(2)   & 295.35(30) \\
324.70(10) & 324.651(25)& 324.532(54)\\
325.80(10) & 325.789(4) & 325.63(76) \\
335.46(16) & 335.4 (2)  & 335.297(54)\\
404.8(3)   & 404.814(4) & 404.849(54)\\
485.00(20) & 485.0(2)   & Observed   \\
503.3(3)   & 503.004(4) & 502.91(12) \\
566.4(7)   & 565.2(2)   & 565.15(12) \\
619.60(20) & 619.60(20) & 619.75(14) \\
637.4(3)   & 636.989(4) & 637.13(10) \\
642.8(7)   & 642.719(5) & 642.46(30) \\
699.80(20) & 699.80(20) & 699.683(77)\\
723.29(14) & 722.911(5) & 722.84(17) \\ \hline
\bottomrule
\end{tabular*}
\end{table}

\begin{table}[H!]
\caption{Primary transitions from the capture state are given. All energies are in keV.}\label{PrimaryGammas}
\begin{tabular*}{\tblwidth}{@{}LLL@{}}
\toprule $E_\gamma$ & $E_f$& $I_\gamma$ (relative)\\ 
\midrule
6524.59(37)  & 80.1854  & \textless{}48.6 \\
6199.78(18)  & 404.815  & 94.8(81)          \\
5904.63(32)  & 699.90   & 100(26)          \\
5691.14(72)  & 913.84   & 25.3(81)       \\
5318.61(13)  & 1285.98  & \textless{}33.8 \\
5085.37(26)  & 1519.404 & 33(11)       \\
4640.48(11)  & 1963.926 & \textless{}48.7 \\
4626.73(39)  & 1977.850 & \textless{}64.4 \\
4623.141(78) & 1981.157 & 20.0(71)        \\
4576.783(54) & 2027.505 & \textless{}73.0 \\
4556.38(11)  & 2048.013 & 43(15)       \\
4200.58(31)  & 2403.92  & \textless{}57.5 \\ \hline
\bottomrule
\end{tabular*}
\end{table}

\begin{figure}%[]
  \centering
    \includegraphics[width=0.5\linewidth]{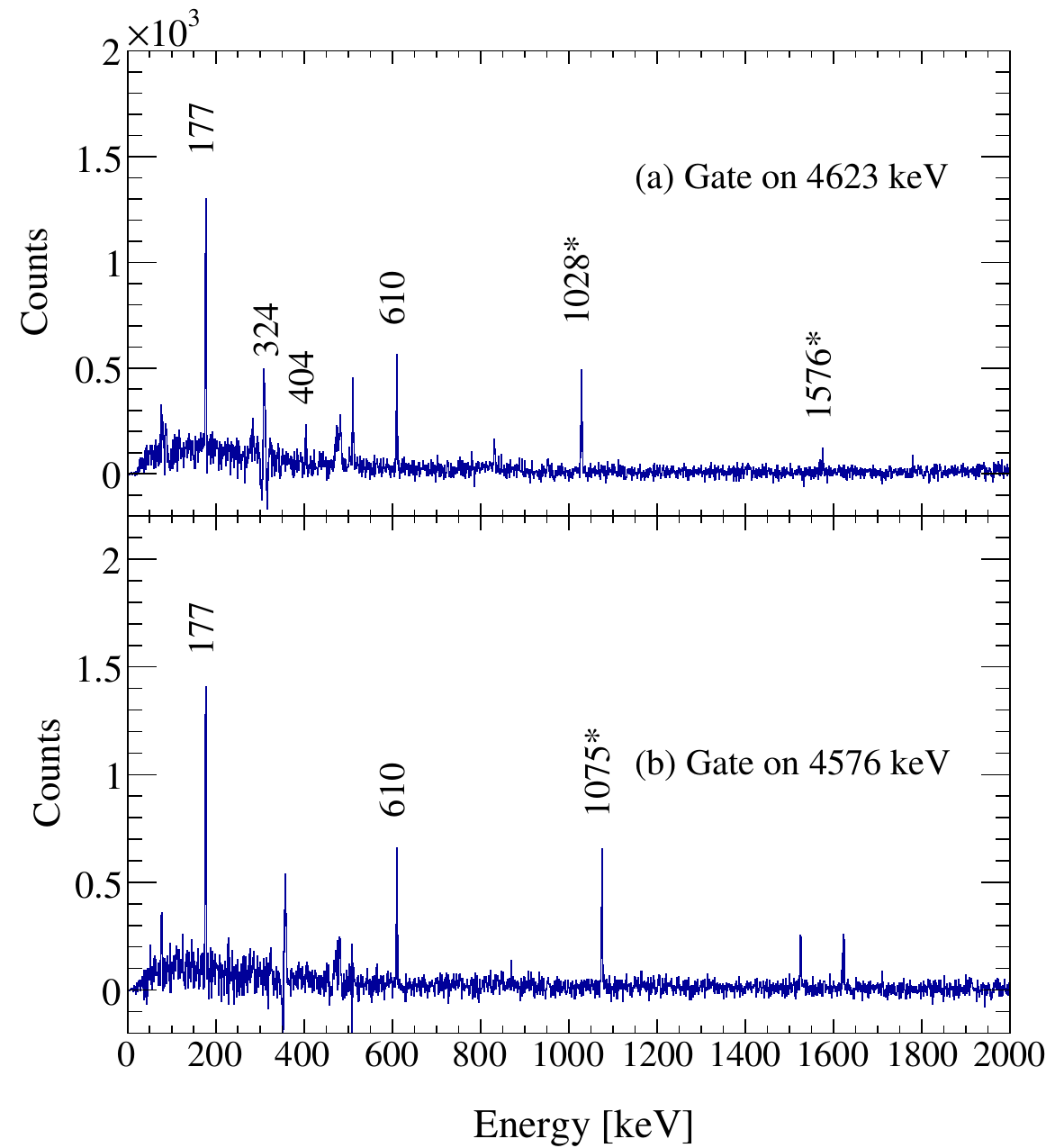}
    \caption{Gates on two of the strongest primary transitions to (a) the 1981 state and (b) the 2027 keV state. Gammas determined to be from \textsuperscript{131}Xe are labelled and those found first by this work are denoted by a $\ast$.}\label{PrimaryGates}
\end{figure}

\begin{figure}%[]
  \centering
    \includegraphics[width=0.5\linewidth]{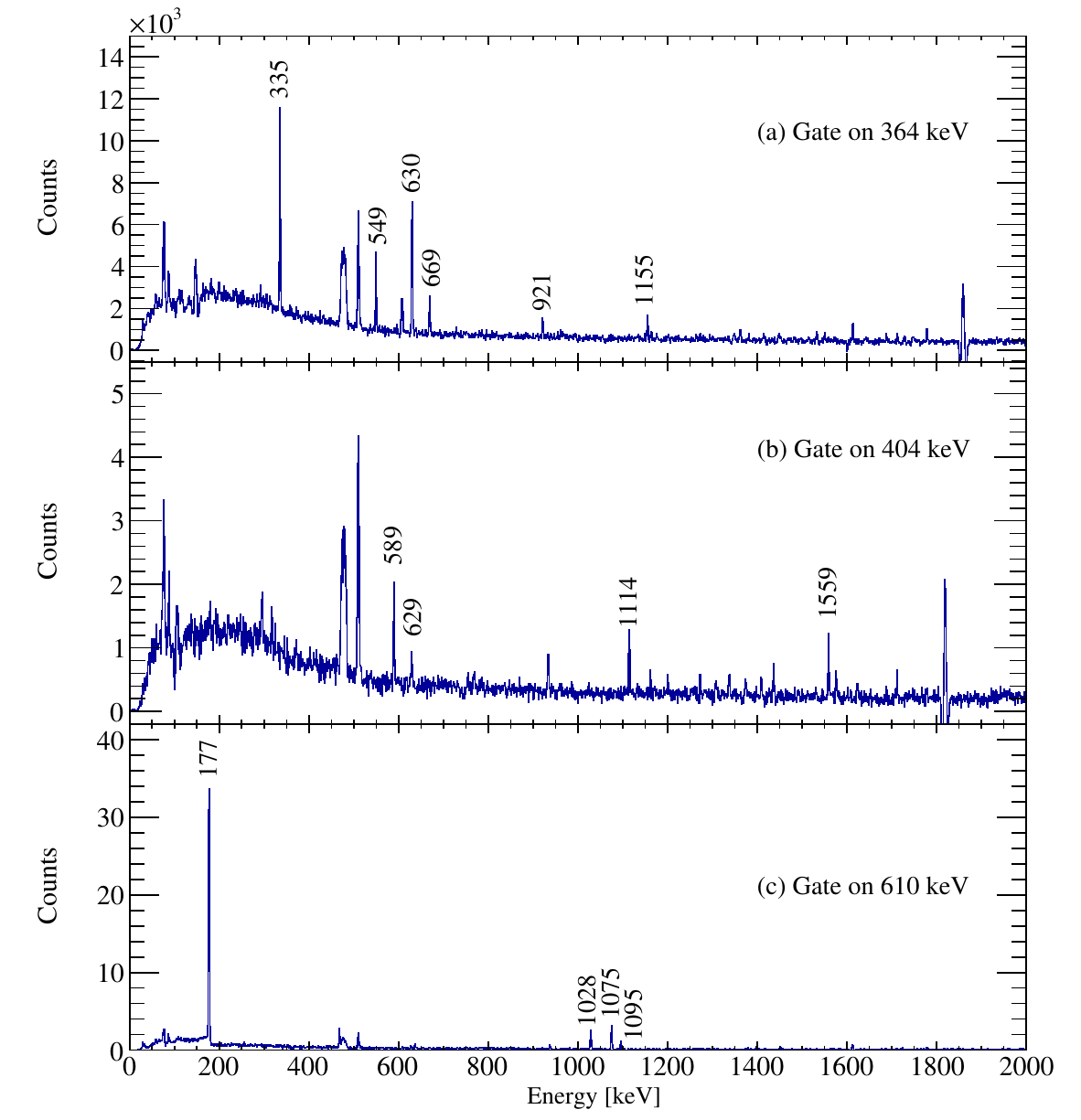}
    \caption{Gates on three previously known gammas are shown to be in coincidence with gammas found by this work that correspond to the existence of new states.}\label{ThreeGates}
\end{figure}

\begin{figure}[h]
  \centering
    \includegraphics[width=1\linewidth]{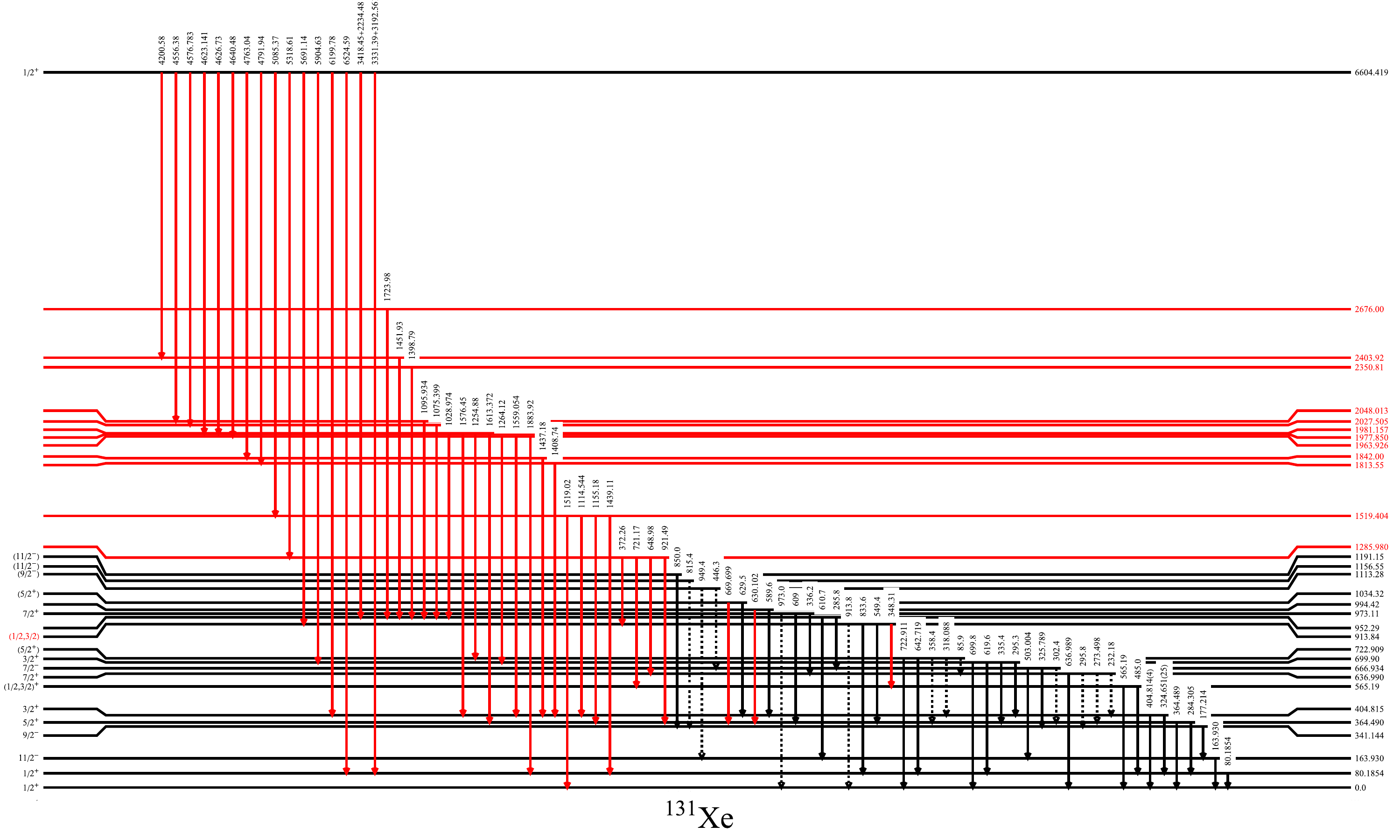}
    \caption{The level scheme of the \textsuperscript{130}Xe(n,$\gamma$)\textsuperscript{131}Xe reaction observed in this work. Previously known states and gammas that were observed are shown in solid black and those not observed in dashed black. New states and gammas found by this work are shown in red and those suspected states as outlined in the above analysis in orange. All energies are in keV.}\label{131XeLevelScheme}
\end{figure}

%ISOTONE FIGURE
\begin{figure}[h]
  \centering
    \includegraphics[width=0.75\linewidth]{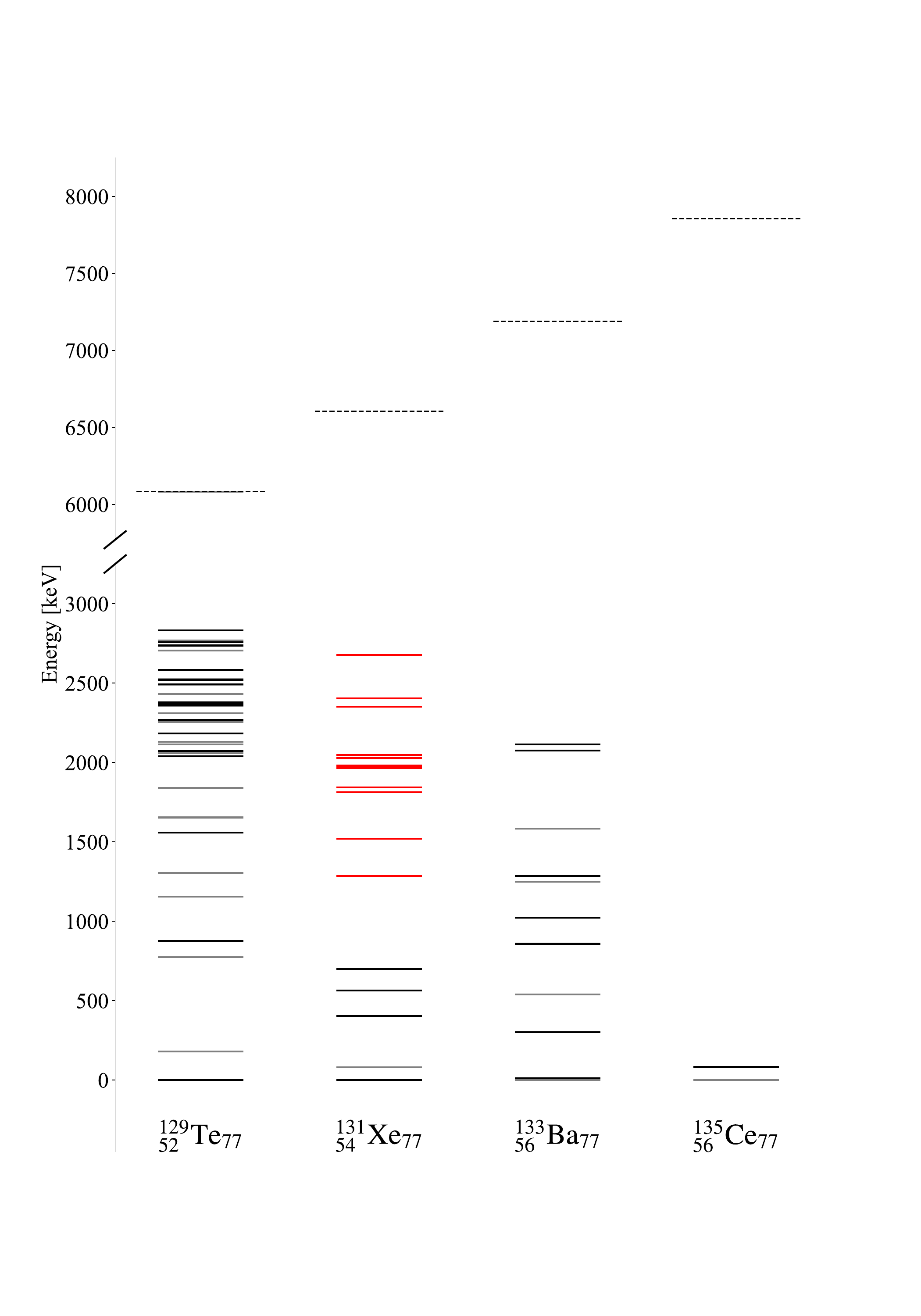}
    \caption{The excited states of spin 1/2 (gray) and 3/2 (black) of isotones of \textsuperscript{131}Xe are shown in comparison with the neutron separation energy (dashed). States of \textsuperscript{131}Xe found by this work are shown in red.}\label{131XeIsotones}
\end{figure}
The analysis of of new transitions and states was done using prompt gamma coincidences: events were reconstructed with a prompt coincidence window of 280 ns. Two matrices were built with these events: $E_\gamma$-$E_\gamma$ ($\gamma\gamma)$, which groups events by individual gamma energy within the coincidence window, and $E_\gamma$-$E_\text{tot}$, which groups events by total energy of all gammas in an event and that of a single gamma. Using $\gamma\gamma$-coincidences, 71 prompt gammas were measured. The resulting spectra from gating on primary transitions is shown in Figure \ref{PrimaryGates} and gating on known previously known transitions in Figure \ref{ThreeGates}. This demonstrates the effectiveness of $\gamma\gamma$-coincidences: though few of these gammas are present in the whole spectrum (Figure \ref{FullSpec}), the gated spectrum greatly reduces background to make otherwise background-buried gammas measurable. All levels seen by this work and their associated gammas are shown in the level scheme in Figure \ref{131XeLevelScheme}. 
\\
\indent These 71 gammas consist of 43 new gammas, of which 12 are primary transitions from the capture state, listed in Table \ref{PrimaryGammas}, and three new transitions between known states, listed and compared in Table \ref{KnownGammas}. A total of 14 new states were found and are enumerated in Table \ref{NewLevels} with the associated gammas and relative intensities. From the primary transitions and known energy levels, the q-value of the reaction was calculated to be 6604.430(54) keV, which is in agreement with the tabulated value of 6604.419(5) keV \cite{nndc_qcalc}. The relative intensities of the primary transitions were determined by gating on the q-value in the $E_{\text{tot}}\text{-}E_\gamma$ spectrum. The branching ratios were determined by gating on transitions that cascade down to a state and determining the counts of each gamma from that state while also correcting for efficiency.
\\
\indent Comparison of previously known \textsuperscript{131}Xe states and gammas to the ones found by this work show that the level density and distribution of gammas is closer to the isotone produced by \textsuperscript{128}Te(n,$\gamma$)\textsuperscript{129}Te \cite{wirth2003} up to around 3 MeV. Figure \ref{131XeIsotones} demonstrates this similarity in level density of 1/2 and 3/2 states. A direct comparison to the isotones produced by \textsuperscript{132}Ba$(\text{n},\gamma)$\textsuperscript{133}Ba \cite{Suliman2009}\cite{PhysRevC.13.2501} and the levels of \textsuperscript{135}Ce \cite{IzvAkadNaukSSSR1972} cannot be done due lack of neutron capture studies and fewer known states and transitions.
In \textsuperscript{131}Xe, based on the presence of primary transitions to all previously known states that are 1/2 or 3/2, it is likely that the spin of the new states that have primary transitions found by this work (Table \ref{NewLevels}) are also either 1/2 or 3/2. This structure is further supported by the high density of nuclear states that are of spin 1/2 or 3/2 in the isotone \textsuperscript{129}Te.

\section{Future Works}\label{FutureWorks}
There are four gammas that were observed to be in coincidence with the resulting cascades from known energy levels but were not observed to have any other transitions from the capture state, known states, or states proposed by this work, nor could they reasonably be attributed to contaminants in the beam. As such, the gammas are reported with the associated energy levels in Table \ref{NewLevels}. It is expected that if additional statistics were available in this experiment, low intensity transitions would be found to the associated energy levels.
Using similar experimental methods, it is the hope of this group to probe the nuclear structure of other xenon isotopes at FIPPS to conduct a longitudinal study of those levels across a single element.

\section{Acknowledgments}\label{Acknowledgments}
Thank you to Milan Krticka for his helpful and enlightening discussion. Thank you to Zsolt R\'evay for his insights on previous measurements of natural xenon prompt gammas collected at Budapest Research Reactor as well as his help in drawing insights from the IAEA Database for Prompt Gamma-ray Neutron Activation Analysis. Thank you to Joseph Lapka for flame sealing the high-purity quartz sample.

J.M. acknowledges funding from the University of Dallas Donald A. Cowan Physics Institute while A.R. and W.F. were supported by the State of Texas through the Texas Nuclear Digital Twin Program at The University of Texas at Austin. 

The raw experimental data and associated metadata are governed by the  Institut Laue-Langevin (ILL) data policy available at \url{https://www.ill.eu}.

\begin{table}
\begin{threeparttable}
\caption{The measured values for the energy levels, gammas, and relative intensities of transitions are compared to previously tabulated values. New gammas and suspected spins are also given. All energies are in keV.}
\label{KnownGammas}
\begin{tabular*}{\tblwidth}{@{}LLLLLLLL@{}}
\toprule $E_i$ (NNDC) & $E_i$ & $\text{J}_i^\pi$ & $E_\gamma$ (NNDC) & $E_\gamma$ & $E_f$ & $I_\gamma$ (relative) & $I_\gamma$ (relative) (NNDC) \\ \midrule 
80.1854(19) & 80.053(54)   & 1/2+       & 80.185(2)   & 80.053(54)   & 0.0     & 100       & 100           \\
163.930(8)  & 163.91(16)   & 11/2$-$    & 163.930(8)  & 163.91(16)   & 0.0     & 100       & 100           \\
341.144(9)  & 341.026(54)  & 9/2$-$     & 177.214(2)  & 177.096(54)  & 163.93  & 100       & 100           \\
364.490(4)  & 364.493(54)  & 5/2+       & 284.305(5)  & 284.089(72)  & 0.0     & 9.1(2.8)  & 7.53(6)       \\
            &              &            & 364.489(5)  & 364.493(54)  & 80.1854 & 100(3)    & 100.0(7)      \\
404.815(4)  & 404.849(54)  & 3/2+       & 324.651(25) & 324.508(54)  & 80.1854 & 49.4(9.3) & 39(5)         \\
            &              &            & 404.814(4)  & 404.849(54)  & 0.0     & 100(7)    & 100(3)        \\
565.19(15)  & 565.15(12)  & (1/2,3/2)+ & 485.0(2)    & Observed\tnote{1}    & 80.1854 &           & 8.7(22)       \\
            &              &            & 565.19(15)  & 565.15(12)  & 0.0     &           & 100.0(22)     \\
636.990(4)  & 637.105(54)  & 7/2+       & 232.18(15)  & Not Observed & 404.815 &           & 0.044(6)      \\
            &              &            & 272.498(17) & Not Observed & 364.490 &           & 0.805(15)     \\
            &              &            & 295.8(2)    & Not Observed & 341.144 &           & 0.025(12)     \\
            &              &            & 636.989(4)  & 637.13(10)  & 0.0     &           & 100.0(13)     \\
666.934(9)  & 666.84(12)   & 7/2$-$     & 302.4(2)    & Not Observed & 364.49  &           & 1.32(16)      \\
            &              &            & 325.789(4)  & 325.63(76)   & 341.144 & 64(15)    & 76(6)         \\
            &              &            & 503.004(4)  & 502.91(12)   & 163.930 & 100(11)   & 100.0(9)      \\
699.90(10)  & 699.997(54)  & 3/2+       & 295.3(2)    & 295.35(30)   & 404.815 & 11(100)   & 13(7)         \\
            &              &            & 335.4(2)    & 335.297(54)  & 364.49  & 94(24)    & 59(11)        \\
            &              &            & 619.6(2)    & 619.75(14)  & 80.1854 & 41(19)    & 46(11)        \\
            &              &            & 699.8(2)    & 699.683(77)  & 0.0     & 100(14)   & 100(16)       \\
722.909(4)  & 722.79(15)   & 5/2+       & 85.9(2)     & Not Observed & 636.990 &           & 0.005(3)      \\
            &              &            & 318.088(16) & Not Observed & 404.815 &           & 4.38(10)      \\
            &              &            & 358.4(2)    & Not Observed & 364.490 &           & 0.9(4)        \\
            &              &            & 642.719(5)  & 642.46(30)   & 80.1854 &           & 12.26(23)     \\
            &              &            & 722.911(5)  & 722.84(17)   & 0.0     &           & 100.0(14)     \\
913.84(14)  & 913.690(54)  &            & 549.4(2)    & 549.207(54)  & 364.49  & 48(21)    & 80(15)        \\
            &              &            & 833.6(2)    & 833.536(70)  & 80.1854 & 100(20)   & 100(8)        \\
            &              &            & 913.8(4)    & 913.42(16)   & 0.0     &           & 2.0(15)       \\
            &              &            &             & 348.67(10)\tnote{2}  & 565.19  &           &               \\
952.29(15)  & 952.058(54)  &            & 285.8(2)    & 285.22(14)   & 666.934 & 15.7(4.7) & 7.7(23)       \\
            &              &            & 610.7(2)    & 610.881(54)  & 341.144 & 100(3)    & 100(16)       \\
971.22(13)  &              & (9/2)+     & 334.3(2)    & Not Observed & 636.99  &           & 5.6(14)       \\
            &              &            & 606.6(2)    & Not Observed & 364.490 &           & 100(12)       \\
973.11(14)  & 973.08(19)   & 7/2+       & 336.2(2)    & 336.09(19)   & 636.990 &           & \textless{}43 \\
            &              &            & 609         & Observed\tnote{3}    & 364.490 &           & 90(3)         \\
            &              &            & 973.0(2)    & Not Observed & 0.0     &           & 100(3)        \\
994.42(20)  & 994.593(54)  &            & 589.6(2)    & 589.783(54)  & 404.815 &           & 100           \\
            &              &            &             & 630.102(54)\tnote{2} & 364.490 &           &               \\
1034.32(20) & 1034.207(54) & (5/2)+     & 629.5(2)    & 629.73(28)   & 404.815 &           & 100           \\
            &              &            &             & 669.699(69)\tnote{2} & 364.490 &           &               \\
1113.28(15) &              & (9/2)-     & 446.3(2)    & Not Observed & 666.934 &           & 38(13)        \\
            &              &            & 949.4(2)    & Not Observed & 163.930 &           & 100(25)       \\
1156.55(20) &              & (11/2)-    & 815.4(2)    & Not Observed & 341.144 &           & 100           \\
1191.15(20) & 1191.193(54) & (11/2-)    & 850.0(2)    & 850.049(54)  & 341.144 & 100       & 100           \\ \hline
\bottomrule
\end{tabular*}
\begin{tablenotes}
    \item[1] Boron contamination (477 keV) in the quartz prevented precise measurement of this gamma.
    \item[2] Transitions found by this work
    \item[3] The prominent 610 keV gamma from \textsuperscript{131}Xe prevented precise measurement of this gamma.
\end{tablenotes}
\end{threeparttable}
\end{table}
\begin{table}[th]
\begin{threeparttable}
\caption{The new energy levels and gammas found by this work are given. All energies are in keV.}
\label{NewLevels}
\begin{tabular*}{\tblwidth}{@{}LLLL@{}}
\toprule $E_i$ & $E_\gamma$ & $E_f$ & $I_\gamma$ (relative) \\ \midrule
1285.980(54)   & 921.49(10)   & 364.49  & 100(54)   \\
               & 721.17(14)   & 565.19  & 91(59)    \\
               & 648.98(7)    & 636.99  & 86(56)    \\
               & 372.26(18)   & 913.84  & 80(53)    \\
1519.404(54)   & 1439.11(24)  & 80.1854 & 14(17)    \\
               & 1155.18(11)  & 364.49  & 39(20)    \\
               & 1114.544(54) & 404.815 & 79(29)    \\
               & 1519.02(23)  & 0.0     & 100(28)   \\
1813.55(24)\tnote{1}   & 1408.74(24)  & 404.815 &           \\
1842.00(65)\tnote{1}   & 1437.18(65)  & 404.815 &           \\
1963.926(54)   & 1883.92(16)  & 80.1854 & 100(14)   \\
               & 1559.054(71) & 404.815 & 54(11)    \\
               & 1264.12(14)  & 699.90  & 30(8)     \\
1977.850(54)   & 1613.372(71) & 364.49  & 100(28)   \\
               & 1254.88(23)  & 722.909 & 83(84)    \\
1981.157(54)   & 1028.974(54) & 952.2   & 100(7)    \\
               & 1576.45(30)  & 404.815 & 38.1(4.4) \\
2027.505(73)   & 1075.399(54) & 952.2   &           \\
2048.013(90)   & 1095.934(54) & 952.2   &           \\
2350.81(11)\tnote{1}   & 1398.79(11)  & 952.2   &           \\
2403.92(18)    & 1451.93(18)  & 952.2   &           \\
2676.00(67)\tnote{1}   & 1723.98(67)  & 952.2   &           \\
Order unclear\tnote{2} & 3331.39(30)  & 80.1854 &           \\
               & 3192.561(29) &         &           \\
Order unclear\tnote{2} & 3418.45(23)  & 952.2   &           \\
               & 2234.48(14)  &         &           \\ \hline
\bottomrule
\end{tabular*}
\begin{tablenotes}
    \item[1] No transitions from the capture state were observed.
    \item[2] The order of the cascade involving these levels is unclear due to a lack of transitions to more than one known level.
\end{tablenotes}
\end{threeparttable}
\end{table}

% Use the style of numbering in square brackets.
% If nothing is used, default style will be taken.
%\begin{enumerate}[a)]
%\item 
%\item 
%\item 
%\end{enumerate}  

%\begin{itemize}
%\item 
%\item 
%\item 
%\end{itemize}  

%\begin{description}
%\item[]
%\item[] 
%\item[] 
%\end{description}  

\clearpage %%Remove this from your manuscript

%% Figure
%\begin{figure}%[]
%  \centering
%%    \includegraphics{}
%    \caption{}\label{fig1}
%\end{figure}
%
%
%\begin{table}%[]
%\caption{}\label{tbl1}
%\begin{tabular*}{\tblwidth}{@{}LL@{}}
%\toprule
%  &  \\ % Table header row
%\midrule
% & \\
% & \\
% & \\
% & \\
%\bottomrule
%\end{tabular*}
%\end{table}

% Uncomment and use as the case may be
%\begin{theorem} 
%\end{theorem}

% Uncomment and use as the case may be
%\begin{lemma} 
%\end{lemma}


\printcredits

\bibliographystyle{elsarticle-num}%cas-model2-names}

\bibliography{cas-refs}



\end{document}